    \documentclass[%
    reprint,
    amsmath,amssymb,
    aps,
    floatfix,
    ]{revtex4-2}
    
    \usepackage{graphicx}% Include figure files
    \usepackage{dcolumn}% Align table columns on decimal point
    \usepackage{bm}% bold math
    \usepackage{xcolor} %colored text
    \usepackage{braket}
    
    \usepackage{verbatim} %for commenting out whole paragraph by using \begin{comment} and \end{comment}
    
    \newcommand*{\bq}[0]{\mathbf{q}}

    \DeclareMathOperator{\tr}{tr}
        
\begin{document}
    
    \preprint{APS/123-QED}
    
    \title{ Site independent strong phonon-vacancy scattering in high temperature ceramics ZrB$_2$ and HfB$_2$}
    
    \author{Vrishali Sonar\textit{$^{1,2}$}}
    \author{Rohan Dehankar\textit{$^{1}$}}
    \author{K. P. Vijayalakshmi\textit{$^{3}$}}
    \author{Natalio Mingo\textit{$^{4}$}}
    \author{Ankita Katre\textit{$^{1}$}}
    \email{ankitamkatre@gmail.com \\ $^{\dagger}$First two authors have contributed equally}
    \affiliation{{$^1$ Department of Scientific Computing, Modeling and Simulation, SP Pune University, Pune-411007 India \\
    $^2$ Dept. of Physics, Adam Mickiewicz University, 61-614 Pozn\`an, Poland \\
    $^3$ Vikram Sarabhai Space Centre, Kochuveli, Thiruvananthapuram-695022, India \\
    $^4$ Universit\'e Grenoble Alpes, CEA, LITEN, 17 rue des Martyrs, 38054 Grenoble Cedex 9, France}}
    
    %Use showkeys class option if keyword
                                  %display desired
    
     \begin{abstract}
         Similar effects of metal and boron vacancies on phonon scattering and lattice thermal conductivity ($\kappa_l$) of ZrB$_2$ and HfB$_2$ are reported. These defects challenge the conventional understanding that associates larger impacts to bigger defects. We find the underlying reason to be a strong local perturbation caused by the boron vacancy that substantially changes the interatomic force constants. In contrast, a long ranged but weaker perturbation is seen in the case metal vacancies. We show that these behaviours originate from a mixed metallic and covalent bonding nature in the metal diborides. The thermal transport calculations are performed in a complete \textit{ab initio} framework based on Boltzmann transport equation and density functional theory. Phonon-vacancy scattering is calculated using \textit{ab initio} Green's function approach. Effects of natural isotopes and grain boundaries on $\kappa_l$ are also systematically investigated, however we find an influential role of vacancies to explain large variations seen in the experiments. We further report a two-order of magnitude difference between the amorphous and pure-crystal limits. Our results outline significant material design aspects for these multi-functional high temperature ceramics.
     \end{abstract}

    \maketitle
    
    %\tableofcontents
    
    \section{\label{sec:level1}Introduction\protect\\}    
Transition metal diborides exhibit a unique range of properties. They have excellent mechanical features such as ultra high melting temperature ($>$ 2500$^\circ$C), high flexural strength of the order of few hundreds of MPa, hardness of few tens of GPa, and appreciable oxidation resistance.\cite{Cheng_JPCL2021,intro-hardness-Nature,intro-Hardne-Mohanavel, intro-hardness,Zapata} Furthermore, they show good electrical and thermal transport.\cite{Lawson-k, Zhang2011ThermalAE} Recently, they were investigated for superconductivity as well as thin film applications too.\cite{Shein_PRB2006, Weiss_MJ2020,sc-2D-nanosheet, sc-MgB2-first, sc-ZrB2_HfB2,sc-Bhatia} Thus, metal diborides are  interesting for low to ultra high temperature applications.\cite{Cheng_JPCL2021, application-nuclear2-power-reactor, Lawson2011AbIC} 

    ZrB$_2$ and HfB$_2$ have gained much attention recently due to their higher stability.\cite{Cheng_JPCL2021} They are actively investigated for applications with a prime focus on thermal transport such as in aerospace vehicles, hypersonic flights, and thermal protective coating,\cite{application-space, application-USA-airforce, application-space-vehicle, Zhang_CS2022} demanding a good understanding of thermal transport in these materials. 
    
    Several experimental studies on the thermal conductivity of ZrB$_2$ and HfB$_2$ (collectively referred as MB$_2$ hereafter) are found in literature.\cite{Zhang2011ThermalAE, Zimmermann-Thermophysical-Properties-of-ZrB2, Intro-Thompson} Zhang et.al.\cite{Zhang2011ThermalAE}  measured diffusivity by photothermal radiometry technique and from that determined the total thermal conductivity ($\kappa$) - the sum of electronic ($\kappa_e$) and lattice thermal conductivity($\kappa_l$). However, they found $\kappa_e$ to be the significant contribution to $\kappa$, whereas $\kappa_l$ is estimated $\sim$10~W/m-K at 300~K for both MB$_2$. 
    A similar trend is found in other experiments too, while not much is mentioned about the purity of MB$_2$ samples except the grain sizes that are reported in $\mu$m range.\cite{McClane1, McClane2, Intro-Thompson} Inquisitively, a few experiments have shown deviation from this trend by reporting more than twice the $\kappa_l$ values at room temperature.\cite{Zimmermann-Thermophysical-Properties-of-ZrB2}  From the computational perspective, Lawson \emph{et al.}.\cite{Lawson-k} implemented molecular dynamics approach. They reported a higher $\kappa_l$ of 54~W/m-K for ZrB$_2$ and 70~W/m-K for HfB$_2$. For ZrB$_2$, \emph{ab initio} study by Xiang \emph{et al.}\cite{Xiang_JECS2019} reported even higher $\kappa_l$ of 90~W/m-K at 300~K, whereas Yang \emph{et al.}\cite{Yang_IJHMT2020} found \emph{ab initio} calculated $\kappa_l$ to be 79~W/m-K and 49~W/m-K without and with phonon-electron scattering respectively. Thus, a large variability in $\kappa_l$ results is seen in literature, where the calculations are consistently found higher than the experiments. One reason for this difference could be the presence of native defects that may appear unintentionally during sample synthesis. Recently, studies exploring the stability of metal and boron vacancy defects in transition metal diborides have also been reported,\cite{Shein_PSS2003, Shein_PRB2006, Johannson_JAP2021} which raises the curiosity to understand their effects on $\kappa_l$ of ZrB$_2$ and HfB$_2$. 
    
    In this work, we present a detailed \emph{ab initio} study of the lattice thermal transport in ZrB$_2$ and HfB$_2$. The crucial roles of vacancies on phonon scattering is investigated along with an understanding of the intrinsic phonon scattering as well as the effect of natural isotopes and grain boundaries. Scattering from the vacancies are calculated with \emph{ab initio} Green's function approach based on $\mathbf{T}$-matrix scattering theory.  
    We find that the metal vacancies ($\Box_\text{M}$) and boron vacancies ($\Box_\text{B}$) show surprisingly similar phonon scattering strengths in both ZrB$_2$ and HfB$_2$, thus demonstrating a distinctive characteristic of \emph{'vacancy site independence'} of the thermal transport in these materials. 
   This behaviour is further investigated by examining the vacancy created perturbations and we find a pronounced effect of the smaller $\Box_\text{B}$ than the bigger $\Box_\text{M}$ in both MB$_2$. 
   Our results provide a tangible explanation for the $\kappa_l$ variations seen in literature for ZrB$_2$ and HfB$_2$.\cite{Zimmermann-Thermophysical-Properties-of-ZrB2, synthesis-zhang} Finally, amorphous limit results are discussed that give an estimate of the $\kappa_l$ lower bound. 
        
    \section{\label{sec:level2}Computational METHODOLOGY}
    
\begin{figure}[t]
      \includegraphics[width=1.0\linewidth]{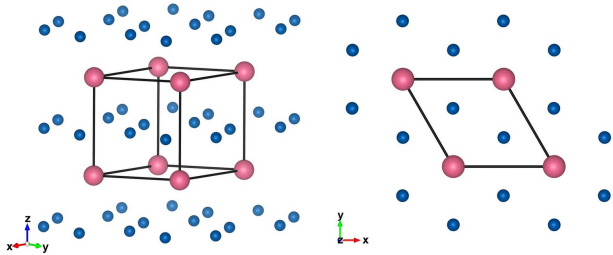}
      \caption{Side and top view of MB$_2$ hexagonal crystal structure (space group $P6/mmm$).  The structure has stacked up alternate layers of metal (pink) and boron (blue) atoms. }
      \label{fig:struct}
    \end{figure}

    \textit{Ab initio} thermal transport calculations are performed using almaBTE code,\cite{alma_CPC2017} with extended implementation of Green's function approach for defects as in Refs.~\cite{AK-SiC, Katre_JMCA2016, AK-GaN}.  The lattice thermal conductivity tensor $\boldsymbol{\kappa_l}$ at a given temperature $T$ is calculated as \cite{alma_CPC2017, ShengBTE_2014, Protik_MTP2017}
    \begin{equation}
    \boldsymbol{\kappa_l} = \frac{1}{k_BT^2N\Omega}\sum_\lambda n_{0_{\lambda}}(n_{0_{\lambda}}+1)(\hbar \omega_\lambda)^2 v_\lambda \otimes v_\lambda\tau_{_\lambda}
    \label{eq:kappa}
    \end{equation}
    
    \noindent where $\omega$ is the phonon frequency, $v$ the group velocity, $n_0$ the equilibrium occupation and $\tau$ the relaxation time for phonon mode $\lambda$. The normalisation factor includes unitcell volume $\Omega$ and $\bq$-point mesh density $N$.  Final $\kappa_l$ is calculated here as an average of in-plane and cross-plane components of thermal conductivity tensor as $\kappa_l = \tr(\boldsymbol{\kappa_l})/3$. Total $\tau$ is given as,
    
    $$ \tau^{-1}_\lambda = \sum_{_{\substack{m=\text{anh,}\\ \text{iso, def, }\hdots}}}\tau^{-1}_{\lambda,m} $$ 
    
    where the summation is over different phonon scattering mechanisms ($m$) for each phonon mode $\lambda$. For perfect infinite lattices of MB$_2$, we include only intrinsic three phonon scattering $\tau^{-1}_{\text{anh}}$ as described in Ref.~\cite{ShengBTE_2014} .Scattering through isotopic mass disorder $\tau^{-1}_{\text{iso}}$ is further included using the methodology by Tamura \emph{et al.} in Refs.\cite{Tamura_1983, Tamura_1984}.
    $\tau^{-1}_{\text{anh}}$ requires \emph{ab initio} calculated third-order interatomic force constants (IFCs) for the crystal, whereas only second-order IFCs and natural abundances of stable isotopes are needed for $\tau^{-1}_{\text{iso}}$.
    
    Phonon scattering rates due to crystallographic point defects, $\tau^{-1}_{\text{def}}$, are calculated using the expression obtained from $\mathbf{T}$-matrix scattering theory and the Optical theorem as,\cite{Mingo_PRB2010, AK-SiC}
    
    \begin{equation}
         \tau^{-1}_{\lambda,\mathrm{def}} = - f_{\mathrm{def}}
        \frac{\Omega}{\Omega_{\mathrm{def}}}\frac{1}{\omega_{\lambda}} \Im\left\lbrace
         \Braket{ \lambda | \mathbf{T} | \lambda }
         \right\rbrace .
      \label{eq:tau_optical}
    \end{equation}
    
    \noindent where $f_{\mathrm{def}}$ is the number fraction of defects and $\Omega_{\mathrm{def}}$ the volume of a defect. The $\mathbf{T}$ matrix is given as, 
    \begin{equation}
        \mathbf{T} = \left(\mathbf{I} - \mathbf{V}\mathbf{g^{+}}\right)^{-1}\mathbf{V} 
      \label{eq:Tmat}
    \end{equation}
    where $\mathbf{g^{+}}$ is the Green's function for perfect crystal, $\mathbf{I}$ the identity matrix and $\mathbf{V}$ the perturbation matrix illustrating the differences between defect and perfect structures. 
    $\mathbf{V}$ is the sum of IFC and mass perturbations, $\mathbf{V}=\mathbf{V}_K + \mathbf{V}_M$, where $\mathbf{V}_M$ is specially the onsite perturbation. In the case of vacancies, this onsite contribution is not relevant due to absence of atom at the defect site. Thus, we have $\mathbf{V} = \mathbf{V}_K$ only for $\Box_{\text{M}}$ and $\Box_{\text{B}}$ in MB$_2$. The IFC perturbations are considered only up to third nearest neighbour of every atom in a larger sphere of radius $\sim$5~\AA\ (6~NN) centred at the vacancy site. $\mathbf{V}_K$ is further corrected using an iterative scheme to annihilate the non-zero terms beyond the cutoff, as discussed in Ref.~\cite{AK-SiC}. 
    
     \begin{figure}[t]
       \includegraphics[width=0.98\linewidth]{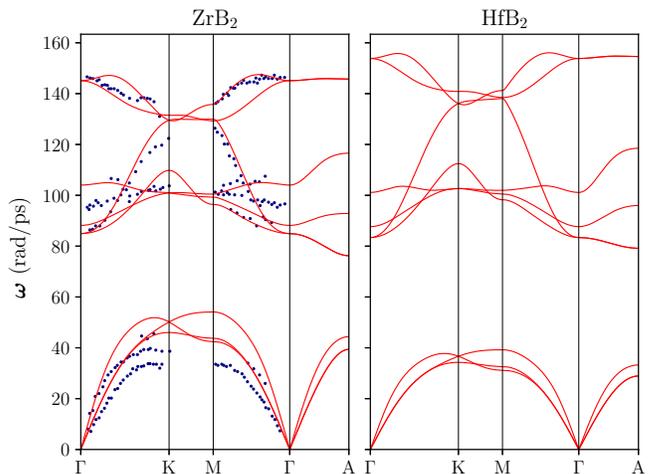}
       \caption{Phonon dispersion of ZrB$_2$ and HfB$_2$. The calculations (lines) are compared with the available experiments (dots) for ZrB$_2$.\cite{dispersion-expt}}
       \label{fig:dispersion}
    \end{figure}

    The atomic forces to determine IFCs for perfect and defect structures are performed using density functional theory (DFT) package Quantum Espresso.\cite{QE-2009, QE-2017, QE-GPU}. 5$\times$5$\times$4 supercells of ZrB$_2$ and HfB$_2$ are considered for IFC calculations. Second- and third-order IFCs are extracted as implemented in Phonopy and thirdorder.py respectively.\cite{phonopy, ShengBTE_2014} Further computational details are found in ESI.

    \section{Results and Discussion}
    \subsection{\label{sec:dispersion} Atomic structure and phonons}
        
    \begin{figure}[t]
       \includegraphics[width=1.0\linewidth]{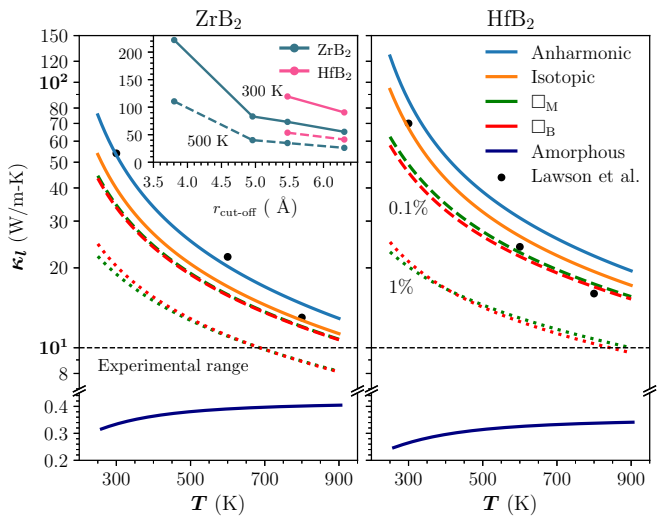}
    \caption{ $\kappa_l$ variation with $T$ for ZrB$_2$ and HfB$_2$ including different phonon scattering contributions.  $0.1~\%$ and $1~\%$ defect concentrations are considered to elucidate the similar effect of $\Box_{\text{M}}$ and $\Box_{\text{B}}$ defects over a range of temperature. Our results are compared with previous calculations by Lawson \emph{et al.}\cite{Lawson-k} and a general experimental range found in most of the literature,\cite{ Zhang2011ThermalAE, Intro-Thompson} (excluding outliers\cite{Zimmermann-Thermophysical-Properties-of-ZrB2}).  Inset shows $\kappa_l$ convergence for ZrB$_2$ and HfB$_2$  with respect to the IFC cut-off radius ($r_{\text{cut-off}}$).}
       \label{fig:kappa}
    \end{figure}
    
    Fig.~\ref{fig:struct} shows the hexagonal crystal structure of MB$_2$ with alternate stacking of metal and boron layers (space group $P6/mmm$). Both the boron and metal layers are planar with covalent and metallic bonding nature respectively. The interlayer bonding in MB$_2$ has mixed ionic and covalent behaviour.\cite{Wang_APLMat2014, Wagner_ZAC2013, Lawson2011AbIC} With such bonding, MB$_2$ has a compact unitcell with three atoms (one metal and two borons) and only slight differences in the $\perp$ and $\parallel$ lattice parameters ($c/a \gtrsim 1$). The obtained relaxed cell parameters for ZrB$_2$ are $a$= 3.17~\AA, $c$=3.54~\AA~and HfB$_2$ are $a$ = 3.13~\AA, $c$=3.47~\AA. 
    
    Phonon dispersion curves for ZrB$_2$ and HfB$_2$ are shown in Fig.\ref{fig:dispersion}. Several overlapping features could be seen for these compounds due to their same crystal structures and the fact that Hf and Zr are iso-group elements. For both MB$_2$, there are in total nine phonon branches at a given $\mathbf{q}$-point (three acoustic and six optic branches) and a similar dispersive nature of these branches is seen in the IBZ path. The differences in the phonon dispersion are mainly seen in the phonon frequency range and acoustic-optic gap. ZrB$_2$ with larger unitcell volume ($\Omega$=31.059~\AA$^{3}$) has slightly narrower frequency range than HfB$_2$ ($\Omega$=29.87~\AA$^{3}$). Bonds are stiffer in HfB$_2$ than ZrB$_2$. Thus, HfB$_2$ bonds can be imagined as stronger springs with larger IFCs, leading to a wider phonon frequency range. Fig.~\ref{fig:dispersion} also shows acoustic-optic energy gaps for both ZrB$_2$ and HfB$_2$, however with different widths. The acoustic-optic (a-o) gap is due to mass variations of the constituting elements (here, $\frac{m_{\text{Zr}}}{m_{\text{B}}}$= 8.44,  $\frac{m_{\text{Hf}}}{m_{\text{B}}}$=16.51), as heavier elements contribute majorly to the low frequency modes and vice-versa ($\omega \propto 1/\sqrt{m}$). Furthermore, as the transition metal mass increases, $m_{\text{Hf}}$ (=174.49 amu) $> m_{\text{Zr}}$ (=91.22 amu), the frequencies of acoustic modes are lowered in HfB$_2$. Overall, we find a good agreement with the experiments produced by electron-energy-loss spectroscopy\cite{dispersion-expt} and previous \textit{ab initio} investigations.\cite{Lawson2011AbIC, phonon-dispersion-Xiang} 
    
    \begin{figure}[t]
       \includegraphics[width=1.0\linewidth]{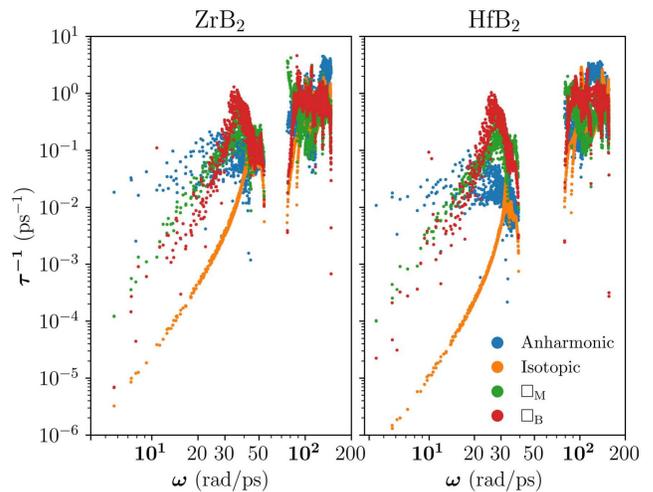}   
       \caption{Phonon scattering contributions from other phonons at 300~K, natural isotopes and vacancy defects in ZrB$_2$ and HfB$_2$ as a function of $\omega$. Vacancy concentration of $1~\%$ is considered here for both $\Box_{\text{M}}$ and $\Box_{\text{B}}$.}
       \label{fig:scattering}
    \end{figure}

    \subsection{\label{sec:tcond} Thermal Conductivity of MB$_2$}
    
    A fair intuition of $\kappa_l$ of a material could be gained from its phonon dispersion. A large a-o phonon gap reduces anharmonic scattering and thus could lead to high thermal conductivity.\cite{AK-GaN, AK-SiC} Similarly, higher phonon group velocities, obtained as $\partial \omega/ \partial \bq$ contribute to a higher $\kappa_l$. However, a further detailed comment requires explicit calculation of the phonon relaxation time $\tau$ and $\kappa_l$. 
    
    Firstly, we take up the pristine cases of ZrB$_2$ and HfB$_2$. Considering a large variation in the literature values,     \cite{Zimmermann-Thermophysical-Properties-of-ZrB2, Zhang2011ThermalAE, Intro-Thompson, Lawson-k, Xiang_JECS2019, kappa-book-Hilmas-Gregory} we perform a convergence analysis of $\kappa_l$ at first. Fig.~\ref{fig:kappa}(inset) shows ZrB$_2$ and HfB$_2$ $\kappa_l$ variation w.r.t. the cut-off radius ($r_{\text{cut-off}}$) used for the IFC calculations. Only within this $r_{\text{cut-off}}$, the double atomic displacements are considered while calculating third order IFCs for $\tau^{-1}_{\text{anh}}$. The analysis is done at two different temperatures and we find convergence for a large $r_{\text{cut-off}}$ of $\sim$6.5~\AA\ (9$^{\text{th}}$ atomic neighbour shell). This analysis helps to avoid the ambiguity of a higher $\kappa_l$ reported previously,\cite{Xiang_JECS2019, Yang_IJHMT2020} that could result from using a randomly chosen small $r_{\text{cut-off}}$. 
        
    The $\kappa_l$ variation with temperature is shown in Fig.~\ref{fig:kappa}, where ZrB$_2$ has lower $\kappa_l$ than HfB$_2$  due to a stronger $\tau^{-1}_{\text{anh}}$ for ZrB$_2$, Fig.~\ref{fig:scattering}. This originates from both (a) smaller a-o gap and (b) larger cell volume of ZrB$_2$. At 300~K, $\kappa_l$ for ZrB$_2$ is nearly half of that of HfB$_2$. We also include the effect isotope scattering - a reportedly significant phonon scattering mechanism for wide a-o gap materials.\cite{Lindsay_PRB2013} $\tau^{-1}_{\text{iso}}$ and $\tau^{-1}_{\text{anh}}$ variation with $\omega$ are shown in Fig.~\ref{fig:scattering}. The expected behaviour of $\tau^{-1}_{\text{iso}}$ prominence at high $\omega$ is seen for MB$_2$, where $\tau^{-1}_{\text{iso}}$ is of the same order of magnitude as $\tau^{-1}_{\text{anh}}$. A substantial effect of the isotope scattering on thermal conductivity is seen over a range of temperature, with a reduction of 25\% and 21\% at 300~K for ZrB$_2$ and HfB$_2$ respectively, arising due to the mass variances ($g$) of mostly the heavier elements in the compounds ($g_{\text{Zr}}$=3.42$\times10^{-4}$, $g_{\text{Hf}}$=5.29$\times10^{-5}$).\cite{Lindsay_PRL2013} The effect of isotope scattering can thus be reduced by synthesizing isotopically enriched samples. Excellent findings on isotopic enrichment have recently been reported for boron nitride.\cite{Chen_Sci2020} 
        
    \subsection{\label{sec:gr} Effective Grain Sizes} 
    
    \begin{figure}[t]
      \includegraphics[width=0.98\linewidth]{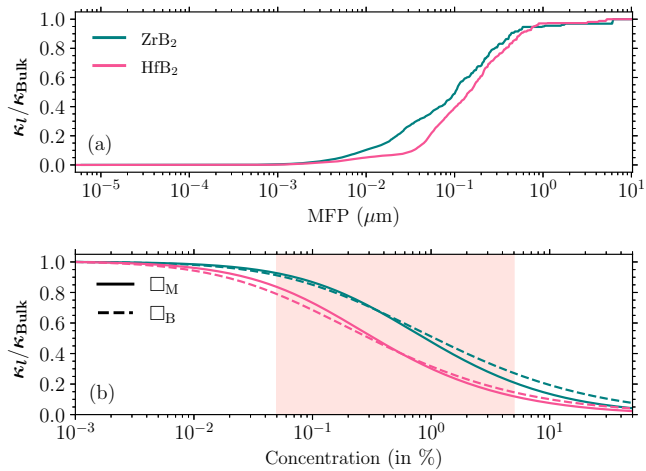} 
      \caption{Variation of $\kappa_l/\kappa_{Bulk}$ at 300~K (a) w.r.t. phonon mean free path (MFP) contributions, and (b) as a function of defect concentration. }
      \label{fig:kl_defconc}
    \end{figure}

    The differences in the experiments and our calculations of $\kappa_l$ are still large, Fig.~\ref{fig:kappa} (labelled 'Isotopic').  Such a difference cannot be explained by only a strong phonon-electron scattering in MB$_2$. Yang \emph{et al.} reported 38$\%$ reduction in $\kappa_l$ of ZrB$_2$ considering phonon-electron interactions, though still higher than the experiments.\cite{Yang_IJHMT2020}  Moreover, this interaction is expected to be weaker in HfB$_2$ due to a wider a-o gap. This hints toward the presence of crystallographic defects in the samples, such as grain boundaries and native point defects, manifesting themselves as strong phonon scatterers. As most of the experiments have reported a low $\kappa_l$,\cite{Zimmermann-Thermophysical-Properties-of-ZrB2, Zhang2011ThermalAE, Intro-Thompson} such defects seem to appear naturally and more frequently during the synthesis of MB$_2$, specially when no other doping or phase mixing is reported. 
    
    Grains smaller than the mean free path (MFP) of the phonons will interfere with phonon propagation, whereas larger grain sizes would have no substantial phonon scattering effect. Previous experiments on ZrB$_2$ and HfB$_2$ have reported grain sizes of the order of a few microns in their samples.\cite{Zimmermann-Thermophysical-Properties-of-ZrB2, Zhang2011ThermalAE} To gauge their effectiveness in reducing the $\kappa_l$, we calculated the mode-wise cumulative contribution of $\kappa_l$ at 300~K as a function of phonon MFP, Fig.~\ref{fig:kl_defconc}(a). We find that $\sim$85~$\%$ of the $\kappa_l$ contributions are from the phonons of MFP=10~nm-1~$\mu$m for both ZrB$_2$ and HfB$_2$. Thus, grain sizes $> 1 \mu$m are unlikely to strongly scatter phonons to witness a very small $\kappa_l$ seen in the experiments.

    \subsection{\label{sec:evac} Effect of Vacancies}
    
    \begin{figure}[t]
      \includegraphics[width=1.0\linewidth]{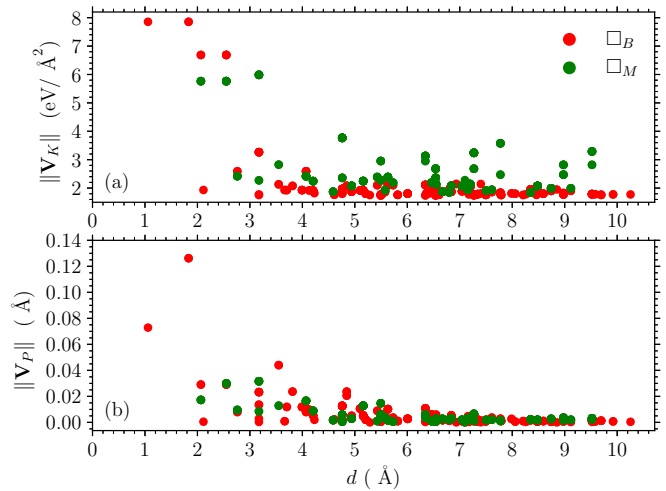} 
      \caption{Squared-$L_2$ norm of the perturbations in IFCs ($\mathbf{V}_K$) and atomic positions ($\mathbf{V}_P$) with reference to the distance from the vacancy ($d$) for $\Box_\text{M}$ and $\Box_\text{B}$ vacancies respectively in ZrB$_2$. Strong, short ranged IFC perturbations are seen for $\Box_\text{B}$, in contrast to weaker, long ranged perturbations for $\Box_\text{M}$. Similar trends are obtained for HfB$_2$, seen in ESI.}
      \label{fig:perturb}
    \end{figure}

    Native defects such as elemental vacancies, antisites and interstitials could be inherent during material synthesis. 
    Out of these, the antisite and interstitial defects in MB$_2$ are found to be energetically unfavourable.\cite{Johannson_JAP2021, Lawson2011AbIC} In fact, our preliminary phonon calculations for B$_{\text{Zr}}$ antisites show dynamical instability (see ESI). Metal and boron vacancies, on the other hand,  have been reported to be present in some concentrations in MB$_2$.\cite{Zhang_PCCP2020, Matas_JPCM2021, Greenwood_QRCS1966, Shein_PSS2003, Dahlqvist_JPCM2015, Johannson_JAP2021} 
      
    Defects change the local environment of charge distribution, mass density, chemical bonding which are static properties but affect the forces seen by dynamic properties. Vacancy defects exhibit their effect through change in these properties only and hence are phenomenologically treated like any other point defect. However, vacancies could create relatively larger distortion in the lattice than other point defects due to the absence of atom and consequently broken local bonds. These distortions act as strong scattering centres for phonons.\cite{Katcho_PRB2014, AK-GaN} The $\tau^{-1}_{\Box_{\text{M}}}$ and $\tau^{-1}_{\Box_{\text{B}}}$ are calculated using Eq.~\ref{eq:tau_optical} for  given vacancy concentrations, Fig.~\ref{fig:scattering}, to further compute $\kappa_l$. Fig.~\ref{fig:kappa} shows the variation of $\kappa_l$ with temperature for only two concentrations - a low ($0.1\%$) and a high ($1\%$) concentration respectively, whereas $\kappa_l$ as a function of vacancy concentration at $T$=300~K is presented in Fig.~\ref{fig:kl_defconc}. Noticeable effects of both $\Box_{\text{M}}$ and $\Box_{\text{B}}$ on $\kappa_l$ are seen, Table.~\ref{table:compare-k} and Fig.~\ref{fig:kappa}. For ZrB$_2$, $\kappa_l$ reduction (at 300~K) of $13\%$ and $15\%$ is seen for 0.1$\%$ of $\Box_{\text{M}}$ and $\Box_{\text{B}}$ respectively, which increases up to $52\%$ and $49\%$ for 1$\%$ defects. Similar trend is seen for HfB$_2$ with $\kappa_l$ reduction of $27\%$ and $32\%$ for $0.1\%$ and $70\%$ and $69\%$ for 1$\%$ of $\Box_{\text{M}}$ and $\Box_{\text{B}}$ respectively. 
        
    One of the striking features seen in our results is the similar effect of $\Box_{\text{M}}$ and $\Box_{\text{B}}$ on $\kappa_l$. The similarity is seen across a wide range of temperature, Fig.~\ref{fig:kappa}, as well as concentration, Fig.~\ref{fig:kl_defconc}. For a quantitative understanding, we introduce a similarity measure of $\kappa_l$s by calculating the geometric mean of the ratio of $\kappa_{l,\Box_{\text{M}}}$ and $\kappa_{l,\Box_{\text{B}}}$ as $$S_{\kappa_l} =  \left(\prod_{i=0}^{n}\kappa_{l,\Box_{\text{M}}}/\kappa_{l,\Box_{\text{B}}}\right)^{1/n},$$ where index $i$ runs through the temperature (T=250$-$900~K) and concentration (c=10$^{-3}$-$10\%$) individually to give $S_{\kappa_l,T}$ and $S_{\kappa_l,c}$ respectively. $S_{\kappa_l}$ should be $\sim1$ to exhibit a good similarity, which we clearly see in our results with $S_{\kappa_l,T}^{0.1\%}$, $S_{\kappa_l,T}^{1\%}$, $S_{\kappa_l,c}^{\text{300~K}}$ equal to 0.99, 1.02, 1.12 for ZrB$_2$ and 0.96, 0.99, 1.08 for HfB$_2$ respectively. 
     Such a trend is due to comparable phonon scattering rates of $\Box_{\text{M}}$ and $\Box_{\text{B}}$, Fig.~\ref{fig:scattering}. The comparison is further explained by finding a Pearson's correlation coefficient of 0.97 in $\tau^{-1}_{\Box_\text{M}}$ and $\tau^{-1}_{\Box_\text{B}}$ using isotonic regression. 
     This is surprising because Boron is less than half the size of and around an order of magnitude lighter than Zr and Hf. Thus, a common understanding is that the transition metal vacancies scatter phonons more strongly than boron vacancies. This is also true in the case of Klemens model that considers vacancy perturbations as $\mathbf{V}=-m_{\Box}-2\left< \bar{m}\right>$, where $m_{\Box}$ is the mass of vacancy and $\left< \bar{m}\right>$ the average atomic mass in the lattice.\cite{Klemens_PR1960, Klemens_PPSSA1955, Gurunathan_PRA2020} Thus, Klemens model suggests a stronger phonon scattering by $\Box_{\text{M}}$ due to a higher mass perturbation than $\Box_{\text{B}}$. However, the explicit consideration of the IFC perturbations ($\mathbf{V}_K$) in our \emph{ab initio} Green's function approach reveals a prominent effect of $\Box_{\text{B}}$ irrespective of its smaller size. 
        
    \setlength{\tabcolsep}{0.95em} % For horizontal padding
    {\renewcommand{\arraystretch}{1.4} % For the vertical padding
     \begin{table}[b]
    \caption{$\kappa_l$ values at 300~K for ZrB$_2$ and HfB$_2$ from our calculations by gradually including  anharmonic scattering, isotope scattering and defect scattering (1~$\%$ concentration), and within amorphous limit. The results are compared with the experiments,\cite{Zimmermann-Thermophysical-Properties-of-ZrB2, Zhang2011ThermalAE}. Multiple values correspond to different samples in the study with the mentioned averaged grain sizes in bracket. Large variation in experiments is evident.} 
    \centering 
    \begin{tabular}{c c c c } 
     \hline\hline 
     &  & \multicolumn{2}{c}{$\kappa_l$ (W/m-K)~~~~~} \\
     &  & ZrB$_2$   &  HfB$_2$ \\
     \hline
      & Anharmonic & 53.20 & 85.46 \\
      &  Isotopic  &        39.78 & 67.24 \\
      This work  &  $\square_{\text{B}}$ (1~$\%$) & 20.30 & 21.12 \\
      &  $\square_{\text{M}}$ (1~$\%$) & 18.94 & 20.19 \\   
      & Amorphous & 0.33 & 0.26 \\
      \hline 
      Expt. & Ref. 20 & 23 & - \\
      & Ref. 7  & 6.25  &  9.09 (5.5 $\mu$m) \\
      & "       &       & 0.6 (10.7 $\mu$m) \\
    \hline 
    \end{tabular}
    \label{table:compare-k} 
    \end{table}
    } 
    
To investigate deeper, we looked into the perturbations created by both the vacancies in MB$_2$, Fig.\ref{fig:perturb}. The squared-$L_2$ norm of $\mathbf{V}_K$ is plotted as a function of the distance from the vacancy ($d$). $\|\mathbf{V}_K\|$ for $\Box_{\text{B}}$ are stronger in the neighbourhood of the vacancy site than that for $\Box_{\text{M}}$, and the effect drop quickly with $d$. In contrast, $\Box_{\text{M}}$ exhibit a long range effect in its perturbations. Perturbations in atomic positions ($\mathbf{V}_{P}$) are also found to be more for $\Box_\text{B}$, suggesting that the prime cause of strong $\tau^{-1}_{\Box_{\text{B}}}$ stems from large atomic displacements near the vacancy, Fig.\ref{fig:perturb}(b). On the other hand, $\Box_\text{M}$ shows some atomic rearrangements only near the defect site. This extraordinary behaviour in ZrB$_2$ and HfB$_2$ comes from their peculiar nature of bonding. Study of electron localization function(ELF) by Lawson et.al\cite{Lawson2011AbIC} highlights the covalent bonding in B layers and metallic bonding between M layers. Covalent bonding is known to be stronger and directional in space, breaking of which leads to an increased local entropy and strong perturbations than that for the case of metallic bonding. Consequently, we see $\tau^{-1}_{\Box_{\text{M}}} \sim \tau^{-1}_{\Box_{\text{B}}}$ in ZrB$_2$ and HfB$_2$, a phenomenon which could be present in other diborides too. Boron is previously found to be a super scatterer in SiC too exhibiting resonant phonon scattering.\cite{AK-SiC}
     
     Fig.~\ref{fig:kl_defconc} further elucidates the region of maximum impact of $\Box_{\text{M}}$ and $\Box_{\text{B}}$ on $\kappa_l$ at a given temperature. A sharp reduction in $\kappa_l$ is seen for the vacancy concentrations $\sim0.05\%$-$5\%$, a range that has been also studied for stability and superconductivity in MB$_2$.\cite{Zhang_PCCP2020, Matas_JPCM2021, Greenwood_QRCS1966, Shein_PSS2003, Dahlqvist_JPCM2015} $\kappa_l$ flattens for concentrations $<0.01\%$ and $>10\%$, 
     where $\tau^{-1}_{\text{anh}}$ leads in the former case and $\tau^{-1}_{\Box_{\text{M/B}}}$ in the latter. 
   
    \subsection{\label{sec:amorhpous} Amorphous limit }
   Finally, to find the lower bound of $\kappa_l$ in MB$_2$, we have calculated the amorphous limit of $\kappa_l$ using Cahill's model that has previously shown good agreement with experiments.\cite{Cahill_PRB1992, Coloyan_APL2016} This lower limit could give an estimation of $\kappa_l$ at extremely high temperatures. As proposed by Cahill \emph{et al.}, the phonon lifetimes in the case of highly disordered structures are taken to be half of their oscillation periods, $\tau=\pi / \omega$.\cite{Cahill_PRB1992} Using these phonon lifetimes and full phonon dispersion, we calculate the thermal conductivity of ZrB$_2$ and HfB$_2$. The obtained results are shown in Fig.~\ref{fig:kappa}, where $\kappa_l < 0.5$~W/m-K is found for both MB$_2$. In contrast to the bulk case, HfB$_2$ has slightly lower $\kappa_l$ in the amorphous case than ZrB$_2$, which is primarily because of higher phonon velocities in ZrB$_2$ (calculated speed of sound - $v_{s_{\text{ZrB}_2}}$ = 7.2~Km/s $ > v_{s_{\text{HfB}_2}}$ = 5.3~Km/s). It is worth noting that amorphous $\kappa_l$ are around two orders of magnitude smaller than the experiments for both MB$_2$, Table.~\ref{table:compare-k}, which indicates that the samples with more disorder and defects could show further low $\kappa_l$ in experiments. This is an important aspect while designing for both low and high thermal conductivity applications of these high temperature ceramics. 
       
    \section{\label{sec:concl} Conclusion}
    
To conclude, we have studied crucial effects of  $\Box_{\text{M}}$ and $\Box_{\text{B}}$ on the $\kappa_l$ of ZrB$_2$ and HfB$_2$. Our results reveal a large effect of smaller boron vacancy having a similar phonon scattering strength as that of bigger metal vacancy. We have explained this behaviour due to strong local perturbation created by boron vacancies. We have also elaborated on other phonon scattering contributions such as isotopes and grain boundaries, where we majorly find that the effective grain sizes for these materials are in the range $<$1~$\mu$m. Furthermore, the amorphous limit of the lattice thermal conductivity is also explored, which serves as the lower bound limit of $\kappa_l$. We find that $\Box_{\text{M}}$ and $\Box_{\text{B}}$ play an important role in explaining the experimentally found small $\kappa_l$s for these diborides. 
    
Our results serve the aim of predictive materials modelling by providing the guidelines to tune the thermal conductivity of ZrB$_2$ and HfB$_2$. Especially, a big control on $\kappa_l$ could be achieved by managing the concentrations of $\Box_{\text{M}}$ and $\Box_{\text{B}}$ in the samples according to the application. In fact, we reveal a surprising prominence of $\Box_{\text{B}}$ in MB$_2$ and a striking behaviour of '\emph{site independent phonon-vacancy scattering}' which could be explored in other materials too having a mixed bonding character. Furthermore, our results for pure crystals (only $\tau^{-1}_{\text{anh}}$) also show that the intrinsic lattice contributions to the thermal conductivity in these ceramics is large, which could be further improved by isotope enrichment. Considering the experimental challenge in tuning the thermal conductivity lies in finding a middle way between the competing density and purity parameters of the samples. With this view, our study provides crucial material design aspects for range of applications for which these metal diborides are actively investigated. 
    
\section{Acknowledgements}
This work is supported by ISRO-RESPOND project (Project No. 190). AK also acknowledges the support from DST-INSPIRE Faculty Scheme (Grant No. IFA17-MS122) and NSM Brahma Super-computing Facility.

    \bibliography{cite}% Produces the bibliography via BibTeX.
    
    \end{document}